\documentclass{aa}
\usepackage{graphicx}
\usepackage{txfonts}
\usepackage{natbib}
\begin{document}
   \title{On the complex X-ray structure tracing the motion of Geminga}

   \author{A.~De Luca
\inst{1}
\and
P.A.~Caraveo
\inst{1}
\and
F.~Mattana
\inst{1,2}
\and
A.~Pellizzoni
\inst{1}
\and
G.F.~Bignami
\inst{3,4,1}
}

\offprints{A. De Luca, deluca@mi.iasf.cnr.it}

\institute{INAF-IASF,  Via Bassini 15, 20133 Milano, Italy
\and
Universit\`a di Milano Bicocca, Dipartimento di Fisica, P.za della Scienza 3, 20126 Milano, Italy 
\and
CESR/CNRS, 9, Av. du colonel Roche, 31028 Toulouse Cedex 4, France
\and
Universit\`a degli Studi di Pavia, Dipartimento di Fisica Nucleare e Teorica, Via Bassi 4, 27100 Pavia, Italy
}

   \date{Received; accepted}

   \abstract{A deep (100 ks) XMM-Newton observation of Geminga has shown two faint
   tails of diffuse X-ray emission, extending for $\sim2'$ behind
   the pulsar, well aligned with the proper motion (PM) direction. 
   We report here on a recent $\sim20$ ks Chandra observation,
   which unveils a new structure,
   $\sim25''$ long and $\sim5''$ thick, starting at the pulsar position and 
   perfectly aligned with the PM direction, with a surface 
   brightness $\sim40$ times higher than that of the XMM Tails.
   The Chandra comet-like feature has a remarkably hard spectrum (photon index 
   $\sim0.9-1.4$) and a luminosity
   of $\sim5.5\times10^{28}$ erg s$^{-1}$, comparable to the energetics of the
   larger XMM one. Geminga is thus the first neutron star to show a clear X-ray evidence 
   of a large-scale, outer bow-shock as well as a
 short, inner cometary trail. 
   \keywords{Stars: neutron -- Pulsars: individual (Geminga) -- X-rays: stars}
   }
\titlerunning{Geminga's X-ray Tails and Trail}
   \maketitle
%

\section{Introduction}
Isolated Neutron Stars (INSs) inherit high space velocities from their supernova explosions.
Moreover, INSs are known to be efficient particle accelerators.
They power a particle wind which is supposed to account for the bulk of their
observed rotational energy loss ($\dot{E}_{rot}$).
When the particle wind from a fast moving INS interacts with the surrounding 
Interstellar Medium (ISM), it
gives rise to complex structures, globally named ``Pulsar Wind Nebulae'' (PWNe)
where $\sim10^{-5}-10^{-3}$ of the INS $\dot{E}_{rot}$ is converted into electromagnetic 
radiation  
\citep[see][for recent reviews]{gaensler04a,gaensler04b}. The study
of PWNe may therefore give insights into the geometry and energetics of the particle wind and,
ultimately, the configuration of the INS magnetosphere
and the mechanisms of particle acceleration. Moreover,
PWNe may probe the surrounding ISM, allowing to measure its density and
ionization state.

A basic classification of PWNe rests on the nature of the  external pressure
confining the neutron star wind \citep{gaensler04b}. For young INSs ($\lesssim$ few 
$10^4$ y) the pressure of the surrounding supernova ejecta plays a major role and a  Crab-like 
PWN is formed. For older systems ($\gtrsim10^5$ y) the INS moves through the unperturbed
ISM and the wind is confined by ram pressure to form a Bow-shock PWN.  

Crab-like PWNe \citep[see][for a review]{slane05} show usually complex morphologies, 
such as tori and/or jets, typically seen in X-rays.
A remarkable axial symmetry,
observed in several cases, is thought to trace the INS rotational axis.
The alignement between the X-ray jets and the INS proper motion directions,
observed for the Crab and Vela pulsars
\citep{caraveo99,caraveo01} implies an alignement between the
rotational axis and the proper motion of the two neutron stars, with possible important 
implications for the understanding of supernova explosion mechanisms \citep{lai01}.

Bow-shocks \citep[see][for reviews]{chatterjee02,gaensler04b}
are observed around older, less energetic INS
and have a simpler,
``velocity-driven'' morphology. They are
seen in $H_{\alpha}$ as arc-shaped structures tracing the forward shock, where
the neutral ISM is suddenly excited. Alternatively, X-ray emission (and/or radio
emission on larger scales) is seen to trail the INS, forming a comet-like
tail, due to synchrotron radiation from the shocked
INS particles diffusing downstream 
\citep[only in the case of PSR B1957+20 both 
the $H_{\alpha}$ and the X-ray structures have been observed,][]{stappers03}.
In X-rays, Bow shocks are typically
fainter than Crab-like PWNe, favouring the detection of nearby objects.

Proximity plays a key role in the case of Geminga.
A deep 100 ks observation with XMM-Newton allowed \citet{caraveo03} to detect
two very faint patterns, or ``Tails'', of diffuse X-ray emission trailing
the neutron star, well aligned along the proper motion direction.

Recently, the Geminga field was imaged by Chandra. 
In the following sections we will report on our analysis of the X-ray data,
including a summary of
the XMM-Newton results. We will then discuss possible
physical interpretations of the complex morphology of the Geminga PWN.

\section{The X-ray data and their analysis}
\subsection{XMM-Newton observation}
   \begin{figure*}[ht!]
   \centering
   \includegraphics[height=14cm,angle=90]{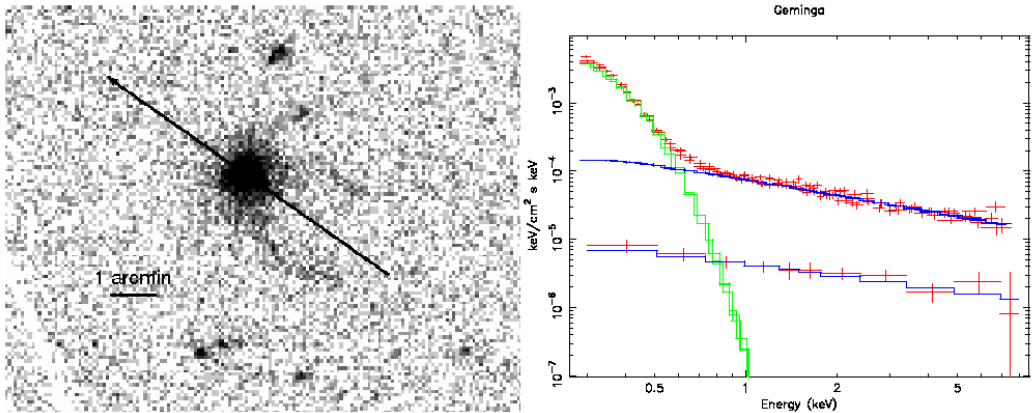} 
   \caption{\label{xmm} The left panel shows the EPIC/MOS image
of the inner Geminga field. Data from MOS1 and MOS2 cameras have been combined. The
presence of two Tails of diffuse emission, aligned with the pulsar proper motion 
direction (marked by the arrow), is apparent. The right panel shows the unfolded spectrum of the 
diffuse Tails (lower plot) compared to the Geminga pulsar unfolded spectrum (upper plot). 
To describe the pulsar spectrum we used 
a double-component model \citep{caraveo04}, encompassing a blackbody 
(T$\sim5\times10^5$ K, green curve), representing thermal 
emission from the star
surface, as well as a power law (photon index $1.7\pm0.1$, blue curve), 
representing non-thermal emission originating in the 
pulsar magnetosphere. The tails' spectrum
is well described by a hard power law (see text).}
    \end{figure*}
Geminga has been imaged on April $4^{th}$, 2002 for 
$\sim100$ ks using the EPIC instrument.
The analysis of the data collected with the MOS cameras
yielded the detection of a 
symmetric structure consisting of two tails $\sim2$ arcmin long trailing the 
pulsar, well aligned with its proper motion vector (see Fig.~\ref{xmm}, left panel).  
The spectrum of both tails, extracted from the $\sim2$ square arcmin region
where the diffuse emission is resolved from the pulsar Point Spread Function
(PSF) wings, is well 
described by a power law with an absorption consistent with that observed 
for the pulsar itself, $N_{\rm H}=1.1\times10^{20}$ cm$^{-2}$ 
\citep[see][and references therein]{caraveo03}.
Using such an $N_{\rm H}$ value, the photon index of the 450-counts tails 
spectrum is $1.6\pm0.2$, similar to the non-thermal component detected in 
the pulsar emission, also shown in Fig.~\ref{xmm} (right panel). 
The observed average surface brightness in the 0.3-5 keV range 
is $\sim10^{-14}$ erg cm$^{-2}$ 
s$^{-1}$ arcmin$^{-2}$; the total unabsorbed 
flux from the region
is $\sim2.1\times10^{-14}$ erg cm$^{-2}$ s$^{-1}$. 
At the 160 pc parallax distance, this translates into a 0.3-5 keV
luminosity of $\sim6.5\times10^{28}$ erg s$^{-1}$, or $\sim2\%$ 
of the total X-ray 
source luminosity in the same range, corresponding to a few 
$10^{-6}$ of Geminga's rotational energy loss.

\subsection{Chandra observation}
The Chandra observation of the field of Geminga started
on 2004, February 7 at 13:02 UT and lasted for 19.9 ks.
The Geminga pulsar was imaged on the back-illuminated
S3 chip of the ACIS detector \citep{burke97}.

In order to reduce the pile-up at the Geminga pulsar position,
the 1/8 subarray mode was used.
The choice of such an operating mode yields
a time resolution of 0.7 s,
but limits the Field Of View (FOV) 
to $\sim1$ arcmin along the CCD readout direction.
   \begin{figure*}
   \centering
\centerline{   \includegraphics[width=8cm,angle=90]{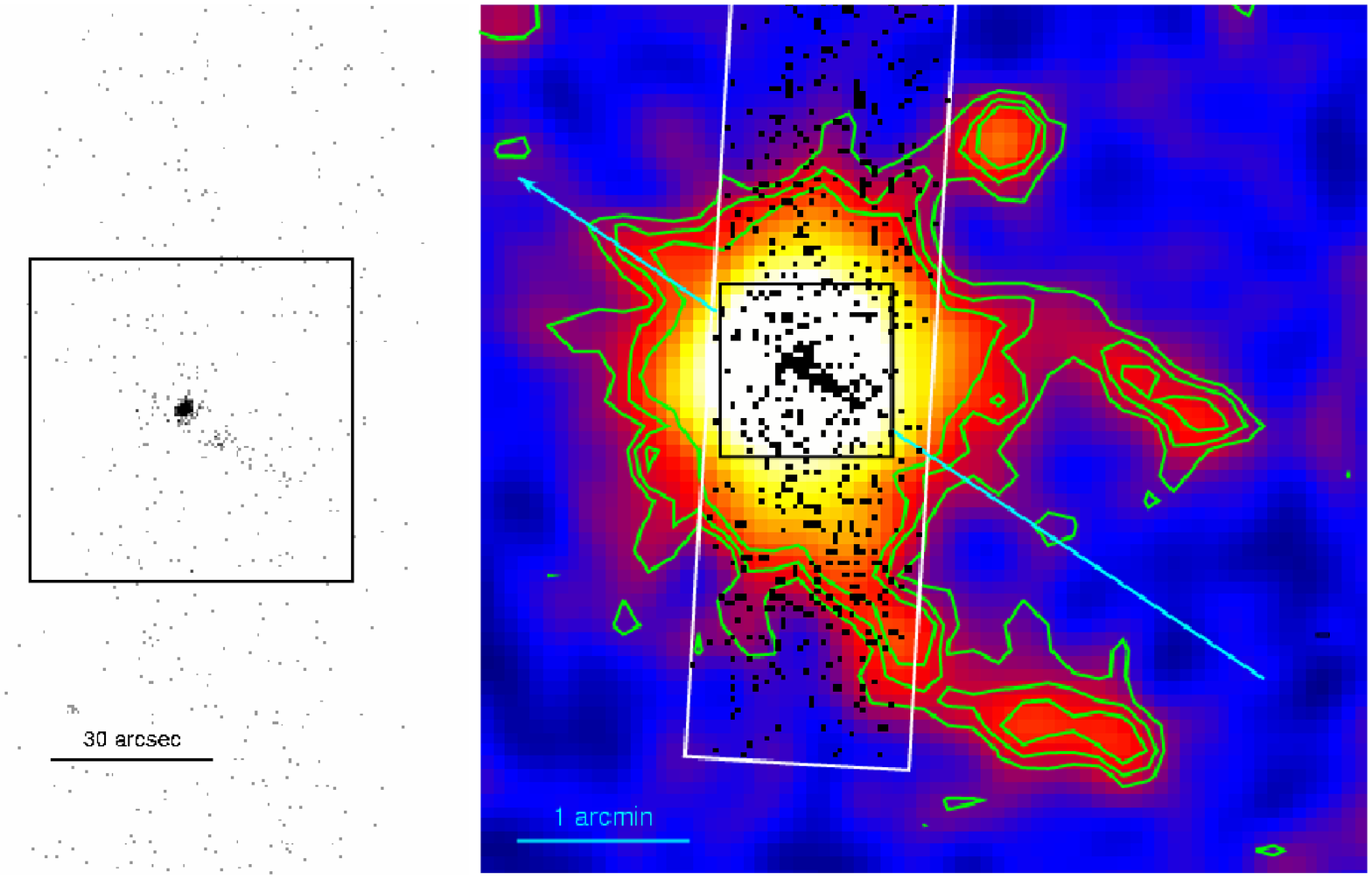} \includegraphics[height=8cm,angle=0]{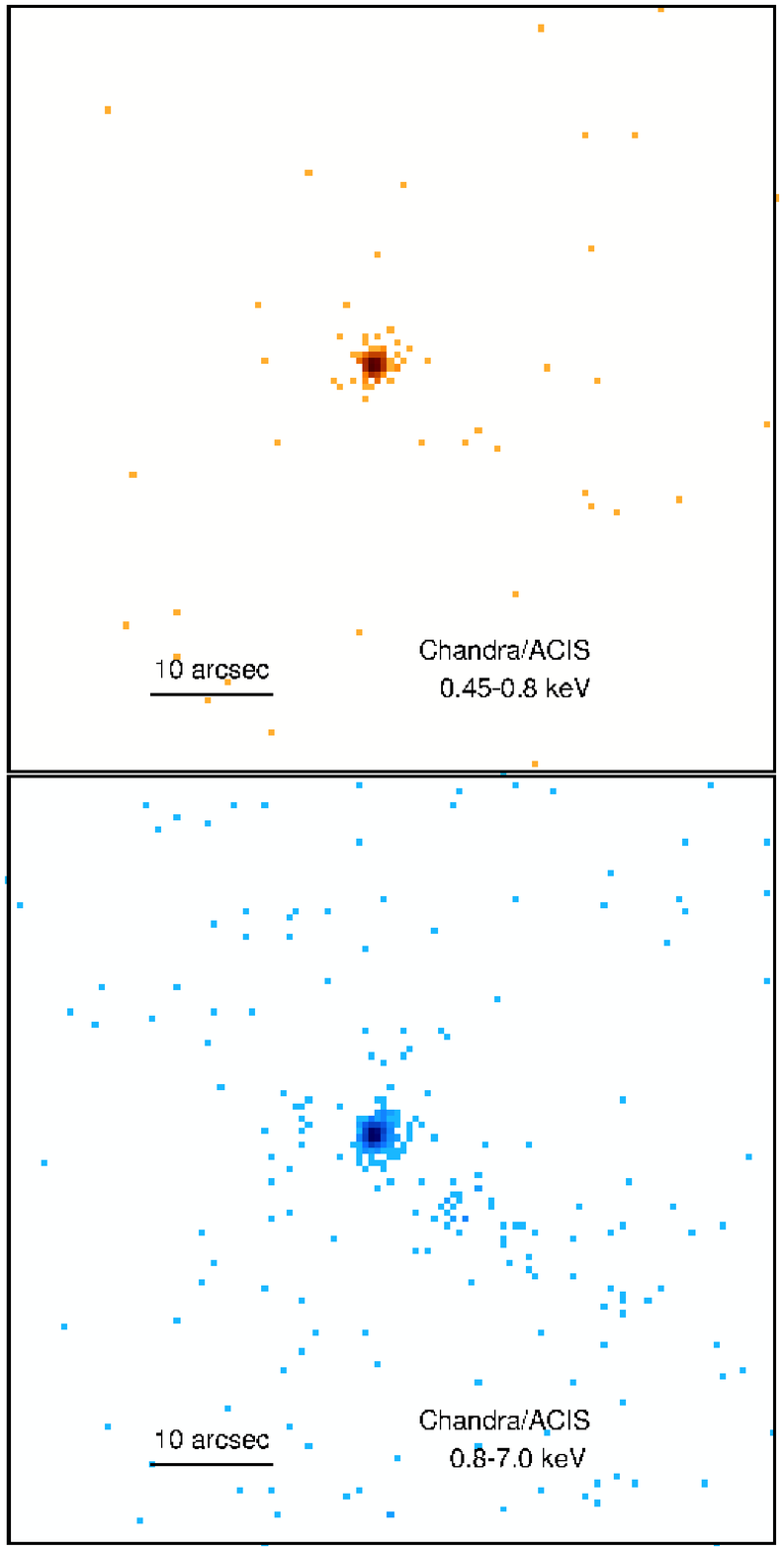} } 
   \caption{\label{chandra} The ACIS image (0.45-7 keV,
0.492$''$ pixel size), zoomed on the Geminga pulsar position, is shown 
in the left panel. 
In the middle panel, the same image,
rebinned to a pixel size of $2''$,  has been superimposed on the
XMM-Newton/MOS smoothed image.
Surface brightness contours for the XMM image have been also plotted. 
The pulsar proper motion direction is marked by
an arrow.
The ACIS field of view is marked by the white rectangular box. The black box identifies
an identical $1'\times1'$ region in all panels.  In the right panel,
the ACIS images in the 0.45-0.8 keV and 0.8-7 keV
ranges are shown. While the Chandra Trail is virtually invisible in the softer band,
it appears very clearly at higher energy, pointing towards a hard spectral shape.}
    \end{figure*}
%

Data were retrieved through the Chandra X-ray Centre (CXC) Archive and
were processed with the CIAO software v.3.2.1, using CALDB v.3.0.1,
to produce calibrated ``level 2'' event lists. No periods of high background
were identified, for a total good exposure time of 18.7 ks.
In our analysis we used only events from the ACIS-S3 chip in the 0.45-7 keV
range.

The resulting ACIS image, zoomed on the Geminga position,
is shown in Fig.~\ref{chandra} (left panel) at its full resolution
(pixel size of $\sim0.5$ arcsec).
In the middle panel of the same figure, the ACIS image,
rebinned to a pixel size of $\sim2''$, has been
superimposed to the XMM-Newton MOS 0.3-8 keV image. 

The Geminga pulsar is clearly detected by Chandra with a count rate of $7.1\pm0.2\times
10^{-2}$ counts s$^{-1}$ (0.45-7 keV, extraction region of $3''$ diameter)
and a spectrum consistent (within the limited
statistics) with the results from XMM-Newton \citep{caraveo04}.
The source coordinates ($\alpha=06^{\rm h} 33^{\rm m} 54.21^{\rm s}$, 
$\delta=17^\circ 46' 14.2''$)
are within $\sim0.7''$
from the values expected on the basis of the source absolute optical
position and proper motion  \citep{caraveo96,caraveo98}.
A faint pattern of diffuse emission, with a
length of $\sim25''$ and a width of $\sim5''$, is also seen to trail the pulsar,
perfectly aligned with the neutron star proper motion direction.
We shall refer to such feature as the ``Chandra Trail'', while the diffuse
emission detected by XMM-Newton will remain the ``XMM Tails''.  
   \begin{figure}
   \centering
   \includegraphics[height=9cm,angle=-90]{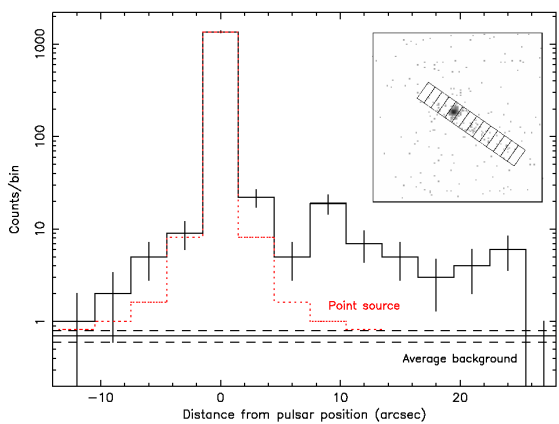}  
   \caption{\label{profile} Brightness  profile of the diffuse Chandra Trail, computed along the 
pulsar proper motion 
direction. Each bin contains the counts detected in a $7''\times3''$ rectangular box, as
shown in the inset. The region represented in the inset is the same marked by the 
black box in Fig.~\ref{chandra}.}
    \end{figure}
%

The brightness profile of the Chandra
Trail along the direction of the pulsar proper motion is shown in
Fig.~\ref{profile}, where each bin corresponds to the counts in a
rectangular area $3''$
long in the PM direction and $7''$ wide perpendicular to it
(see inset of Fig.~\ref{profile}); 
the background level was evaluated from a $60''\times90''$ region
North of the source region, entirely read by the same CCD node.
In order to quantify the contribution from the bright pulsar in the bins
immediately surrounding its centroid,
an ad-hoc PSF was simulated using the {\em ChaRT} software, following
the CXC threads and using the known spectrum of Geminga. The estimated
profile  of the point source within the histogram bins has been plotted in  Fig.~\ref{profile}
as a dotted line. Although a mild pile-up is possibly distorting a little the actual
PSF, Fig.~\ref{profile} suggests that 
no significant diffuse emission is detected ahead of the pulsar. 
Behind the pulsar, 
the Chandra Trail extends along
the proper motion direction up to a distance of $\sim25''$, close to the limit of the 
instrument FOV.
Considering a trail region $22''$ long and $7''$ thick
extending behind Geminga, and neglecting the region within $4.5''$ of
the pulsar, a total (background-subtracted)
of $47\pm7$ counts (in 0.45-7 keV energy range) are detected.

The bulk of the Chandra Trail counts
is found above 0.8 keV. Fig.~\ref{chandra} (right panel) shows the very hard nature
of the Trail emission. Although
the small number of photons hampers a precise charachterization of the Chandra
Trail emission parameters,
we used the maximum likelihood method
implemented in Sherpa with the CSTAT statistic \citep{cash79} to fit the spectra
with an absorbed power law model, fixing the $N_{\rm H}$ 
to the value derived from X-ray fits of the Geminga pulsar spectrum
($N_{\rm H}=1.1\times10^{20}$ cm$^{-2}$).
The model describes well the data (the Q value - namely, the probability to
observe the measured CSTAT statistics or higher if the model is true - is 0.65) and yields
$\Gamma=1.1^{+0.3}_{-0.2}$ ($1\sigma$).
Such a range is 
compatible with (if somewhat harder) the photon index seen for the XMM Tails.
The observed Chandra Trail flux in the 0.45-7 keV range is $\sim2.3\times10^{-14}$ erg cm$^{-2}$ s$^{-1}$.
We note that the Chandra Trail contributes $\sim7\%$ of the Geminga non-thermal flux
as measured by XMM-Newton \citep{caraveo04,deluca05}, 
implying that the actual Geminga pulsed fraction above 2 keV is slightly
higher than previously reported. However, the correction
is smaller than the statistical error on the high energy pulsed fraction quoted by \citet{deluca05}.
In the 0.3-5 keV range,  the Chandra Trail
has a surface brightness of $\sim4\times10^{-13}$ erg cm$^{-2}$ s$^{-1}$ arcmin$^{-2}$,
i.e. $\sim40$ times higher than the XMM Tails' one. 
The unabsorbed flux from the Chandra Trail 
corresponds to a luminosity of $\sim5.2\times10^{28}$ erg s$^{-1}$, which is 
very similar to the estimated XMM Tails' energetic. Thus, the fraction
of the pulsar $\dot{E}_{rot}$ converted into X-ray radiation 
in the arcsec Chandra Trail and in the arcmin XMM Tails is comparable. 

The limited FOV of the Chandra observation hampers a direct comparison
between Chandra and XMM datasets. However it is worth noticing that in the
immediate surrounding of the INS the sensitivity reached by the short Chandra 
observation would not allow the detection of a diffuse emission as faint as the XMM Tails.
Indeed, we estimate the Chandra upper limit for diffuse emission at a level of 
$\sim10^{-13}$ erg cm$^{-2}$ s$^{-1}$ arcmin$^{-2}$
(0.45-7 keV, using a $22''\times7''$ region), corresponding to a flux 10 times 
higher than that reported for the XMM Tails.
Thus, additional diffuse emission with a surface brightness up to
10 times the value detected by XMM-Newton could easily hide in
the Chandra data, making it difficult to obtain a complete picture of the
structures around Geminga.

\section{Discussion}
While the Chandra observation is both too short and with a FOV too limited
to shed light on the nature of the XMM Tails, it has added yet another piece 
to the puzzle of Geminga. A bright, $25''$--long trail of hard X-ray emission 
is now clearly seen behind the pulsar, well aligned with the proper motion 
direction. 

Thus, Geminga joins a small handful of pulsars sporting elongated,
``cometary'' X-ray diffuse structures, such as, e.g., the Mouse 
\citep[PSR J1747-2958,][]{gaensler04a}, 
the Duck \citep[PSR B1757-24,][]{kaspi01,gvaramadze04}, 
the Black Widow \citep[PSR B1957+20,][]{stappers03} and 
PSR B1951+32 \citep{li05} which have all been 
interpreted within a pulsar bow-shock scenario. 

Although similar to the case of PSR B1951+32 \citep[embedded in the complex 
morphology of the SNR CTB 80,][]{li05}, the Geminga X-ray 
emission has a unique symmetry around the pulsar direction of motion.
Indeed, the overall morphology of the diffuse features surrounding
Geminga, with the small Chandra Trail inside the boundaries of the larger, 
arc-shaped XMM Tails, is reminiscent of the composite $H_{\alpha}$/X-ray nebula
associated to the Black Widow. Such nebula is quoted as the most spectacular 
(and so far unique) confirmation of recent bow-shock theory  
\citep[e.g.][]{bucciantini02,vanderswaluw03,gaensler04a,bucciantini05}, 
reflecting the double shock nature of the interaction of a fast moving 
neutron star with the ISM. An outer arc of $H_{\alpha}$ 
radiation is seen, tracing the forward shock, the observed emission coming 
from collisionally excited, shocked ISM; an inner, cometary X-ray feature 
is also seen, tracing the region closer to the pulsar, where the 
shocked pulsar wind emits synchrotron radiation.
The case of Geminga show an obvious and striking difference: both the 
outer, arc-shaped, and the inner, cometary, features are seen in X-rays,
with hard, synchrotron-like spectral shapes. 

The XMM-Newton Tails 
were interpreted as a Bow-shock, traced by very
energetic ($\sim10^{14}$ eV) electrons accelerated by the INS, girating
and emitting synchrotron radiation in the shock compressed ambient 
medium magnetic field \citep{caraveo03}.
Assuming a spherical pulsar wind in a homogeneous ISM and appling a 
simple three-dimensional bow-shock model \citep{wilkin96} to the XMM data 
it was possible to provide an estimate of the unresolved forward
 stand-off angle $\theta_{fw}$, given by the balance between the ram 
pressure and the wind pressure ($\theta_{fw}$$\sim20''-30''$) and to 
constrain the inclination angle of Geminga proper motion with 
respect to the plane of sky (i$<$30$^\circ$) as well as the total 
ISM density (n$\sim$0.1/cm$^3$).
Indeed, 
if $\theta_{fw}\sim20''-30''$ the Chandra Trail would be located in the unshocked pulsar 
wind region, where no strong synchrotron emission is expected \citep[e.g.][]{gaensler04a}.
However, owing to the large uncertainty involved in fitting the bow-shock 
model profile to the XMM diffuse feature (which is unresolved from the bright
pulsar in its whole ``forward'' portion), the actual value of $\theta_{fw}$
could be lower and the Chandra Trail could still fit
in the proposed scenario of pulsar wind confinement. 
A forward stand-off distance $\theta_{fw}$ of $\sim10''$ could be marginally consistent
with both XMM and Chandra data. While implying a revision
of the system parameters (e.g. the density of
the ISM would be $\sim0.5$ cm$^{-3}$), such a scenario would allow to interpret
the Chandra Trail as the surface of the termination shock
(i.e. the inner interface with the unshocked pulsar wind cavity), where the pulsar wind particles 
are shocked and accelerated,
and start emitting synchrotron radiation. The same interpretation was proposed for other examples 
of cometary X-ray nebulae explained within a bow-shock scenario, e.g. the Tongue of the 
Mouse \citep{gaensler04a} and the nebula trailing PSR B1757-24 \citep{gvaramadze04},
which have indeed a similar morphology. 
The hard spectrum of the Chandra trail ($\Gamma \sim 0.9-1.4$) is consistent,
within the large uncertainty,  with
synchrotron radiation by freshly injected electrons in the slow cooling regime
expected by standard shock theory \citep{chevalier00}. 
The overall shape of the termination shock is elongated 
since the ram pressure, confining the pulsar wind in the forward direction, is 
larger than the pressure in the direction opposite to the proper motion. 
As discussed by \citet{gaensler04a}, this is also expected on the basis of
detailed simulations of bow-shock systems 
\citep[e.g.][]{bucciantini02,vanderswaluw03,gaensler04a,bucciantini05}. 

Thus, within such an interpretation the Chandra Trail follows the surface of
the pulsar wind termination shock, while the XMM-Newton Tails trace
the larger scale, limb brightened, interaction of the pulsar wind with the
ISM. Such an interaction, resulting in the formation of a bow-shock, could
take place either through a direct contact leading to a mixing between the INS wind and
the compressed ISM \citep{caraveo03}, or through a contact discontinuity
between the two shocked media, as expected by recent simulations
\citep{bucciantini02,vanderswaluw03,gaensler04a,bucciantini05}. 
In the latter scenario, the XMM tails could be associated 
either to the shocked wind region (where electrons of the pulsar wind
emit synchrotron radiation), or 
to the 
outer region of the shocked ISM, downstream
of the bow-shock interface with the unperturbed ISM
(i.e., the region which is usually associated to $H_{\alpha}$ emission in other cases
of pulsar bow-shocks). 
Hard X-ray emission could be produced there via the synchrotron process 
by energetic electrons of the shocked pulsar wind leaking into such a region, permeated
by the interstellar compressed magnetic field, possibly higher than the magnetic field 
in the shocked wind region. 

Totally different scenarios, where the XMM Tails and the Chandra Trail are unrelated,
could also be explored.
We note that the phenomenology
of the Chandra trail is reminiscent of the jet-like collimated
structures seen in the case of Vela
\citep{helfand01,pavlov03}, another example of nearby INS shining
in high energy $\gamma$-rays.
In particular, the Geminga's Chandra Trail can be
compared to the 
``inner counterjet'' of the Vela PSR
\citep{pavlov03}
in terms of their absolute projected dimensions ($\sim5\times10^{16}$ cm),
spectral shape (photon index $\sim$1.2) and efficiency
($L_{X}$$\sim$$10^{-6} \dot{E}_{rot}$). 
Although the complex features of
the Vela PWN still lack a consistent explanation, and their
3-D morphology and orientation is far from understood \citep{pavlov03}, they
clearly show that the assumption of an isotropic wind outflow is not adequate.
As in the case of Vela, Geminga's diffuse features are characterized by a 
definite symmetry along the proper motion direction. 

Before drawing conclusions on the morphology of the structures
trailing Geminga, we would like to underline that
the combined Chandra/XMM-Newton image shown in Fig.~\ref{chandra}, however tantalizing, does not provide 
a complete and unbiased picture of the INS surroundings.
We see bright emission next to the INS and faint emission far from it but,
owing to instrument characteristics
and/or observing time limitations, Fig.~\ref{chandra} lacks sensitivity in the
intermediate flux region where additional structures may be present.
Only a deep Chandra observation could yield a complete picture of the system 
morphology, especially in its forward portion, the piece of evidence needed to 
ultimately understand its physics.

\begin{acknowledgements}
XMM-Newton and Chandra data analysis is supported by the Italian Space Agency (ASI).
ADL and FM acknowledge an ASI fellowship. We thank an anonymous referee for his/her very 
helpful comments.
\end{acknowledgements}

\end{document}